\documentstyle[11pt,aaspp4]{article}

\begin{document}
\title{Photometry of GX~349+2: Evidence for a 22-hour Period}
\author{Stefanie Wachter\altaffilmark{1} and Bruce Margon\altaffilmark{1}}  
\affil{Department of Astronomy\\
       University of Washington\\
       Box 351580\\
       Seattle, WA 98195-1580\\
       Electronic mail: wachter \& margon@astro.washington.edu}
\altaffiltext{1}{Visiting Astronomer, Cerro Tololo Interamerican Observatory,
National Optical Astronomy Observatories, operated by AURA, Inc. under 
cooperative agreement with the NSF.}
\authoremail{wachter@astro.washington.edu}

\vskip .3in
\begin{center}
Accepted for publication in the Astronomical Journal\\
Volume 112, December 1996
\end{center}

\begin{abstract}

The intense galactic X-ray source GX~349+2 (Sco~X-2) belongs to the class
of persistently bright low-mass X-ray binaries called Z-sources.  
GX~349+2, although observed in X-rays for more than 30 years,   
has only recently been optically identified with a 19th mag star. 
Of the six known Z-sources, only two (\mbox{Sco~X-1} and Cyg~X-2) have 
been studied   
in the optical. It has been suggested that Z-sources as a group are
characterized by evolved companions and correspondingly long orbital  
periods (Sco~X-1, $P=0.8$~d; Cyg~X-2, $P=9.8$~d). Recently 
\markcite{South}Southwell et al.\ (1996) 
have presented spectroscopic observations of GX~349+2
suggesting a 14~d orbital period. We have obtained broadband photometry of   
the system on six consecutive nights, and find a statistically significant 
\mbox{$21.85 \pm 0.4$~h ($3\sigma$)} period of 0.14~mag half-amplitude, 
superposed on erratic flickering
typical of Sco~X-1 type objects. As with other Z-sources, caution will 
be needed to insure that the variations are truly periodic, and not 
simply due to chaotic variability observed over a relatively short time span. 
Depending on the origin of the brightness variations, our proposed 
period could be
either the orbital or half the orbital period.
If our period
is confirmed, then the nature of the 14~d spectroscopic variation found by
\markcite{South}Southwell et al.\ (1996) is unclear. There is evidence that 
the mass function
of GX~349+2 is similar to that of Sco~X-1.

\end{abstract}


\newpage

\section{Introduction}

It has been recognized recently \markcite{HvK}(Hasinger \& van der Klis 1989) 
that 
the brightest low mass X-ray binaries (LMXBs)
fall into two classes characterized by their X-ray time and
spectral variability. They are termed Z-sources and Atoll-sources according to
the pattern they trace in the X-ray color-color diagram.
Currently known Z-sources and their main
characteristics are listed in Table~\ref{t1}.
GX~349+2 (Sco~X-2) belongs to the Z-sources and is one of the 
brightest X-ray sources in the sky.
It is well studied in X-rays 
(\markcite{Sch}Schulz et al. 1989, \markcite{V}Vrtilek et al. 1991, 
\markcite{PCS}Ponman et al. 1988), but
optical data on the system are sparse.
Despite the modest extinction towards GX~349+2, a faint ($V\sim19$)
optical counterpart was not identified until the discovery of its highly
variable radio emission
\markcite{CP}(Cooke \& Ponman 1991) and the
resultant improved positional accuracy.
The one published spectrum of the counterpart
is of both low
resolution and quality \markcite{PA}(Penninx \& Augusteijn 1991). The only 
visible
feature is a strong H$\alpha$
emission line.

\begin{table}
\dummytable\label{t1}
\end{table}
 
The  Z-sources are
believed to have both higher accretion rates and stronger neutron star 
magnetic fields
than the Atoll-sources. This might be caused by a difference in evolutionary
history leading to evolved companions for Z- and dwarf companions for
Atoll-sources \markcite{LP}(Lewin \& van Paradijs 1985), although this 
remains to be proven. Longer periods are 
consequently predicted for Z-sources when
compared to Atoll-sources.
Out of six Z-sources only two have
determined periods, namely Sco~X-1 with 0.79~d and Cyg~X-2
with 9.8~d, while known Atoll-source periods are consistently around 4~h. 
The Z- and Atoll-sources are believed to be the prototypes of LMXBs and are
expected to play a key role in understanding the evolution of all LMXB
systems.

\section{Observations}

\subsection{Optical Photometry}

CCD photometry of GX~349+2 was performed with the CTIO  
0.9~m telescope
from 1995 May 28 UT to June 2 UT. The first three nights were affected by poor
weather and only fairly short (2--3~h) data sets could be obtained. 
During the last three nights, the conditions improved and continuous 
coverage of up to 
8~h was possible. 
The $R$ filter was used through the majority of the
observations.
Overscan and bias corrections were made for each CCD image with the 
task {\it quadproc} at CTIO to deal with the 4 amplifier readout.  
The data were flat-fielded in the standard 
manner with IRAF.
 
\begin{figure}
\plotone{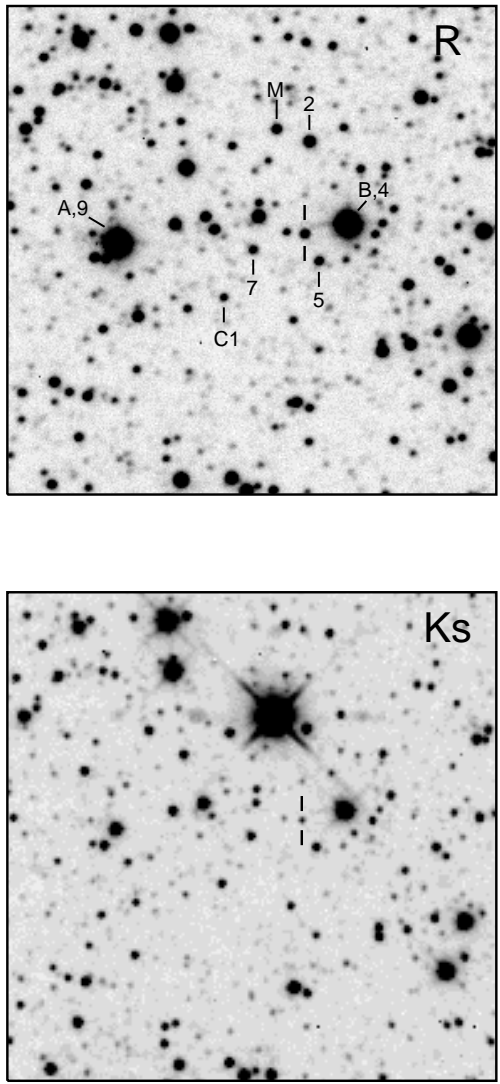}
\figcaption {{\it Top:} 350~s $R$ band exposure of the field of
GX~349+2 taken with the CTIO 0.9~m telescope.  North is up and east is
to the left; the field size is 2\arcmin$\times$2\arcmin.  The
designation of the objects follows Jernigan et al.\ (1978) and Penston
et al.\ (1975) except for objects M and C1 which had no previous
designations. The X-ray counterpart has also previously been called
star 6.  {\it Bottom:} Average of 15 60~s $Ks$ band exposures of the
same field taken with the CTIO 1.5~m telescope. GX~349+2 is marked.
\label{f-fc}}
\end{figure}

\begin{figure}
\plotone{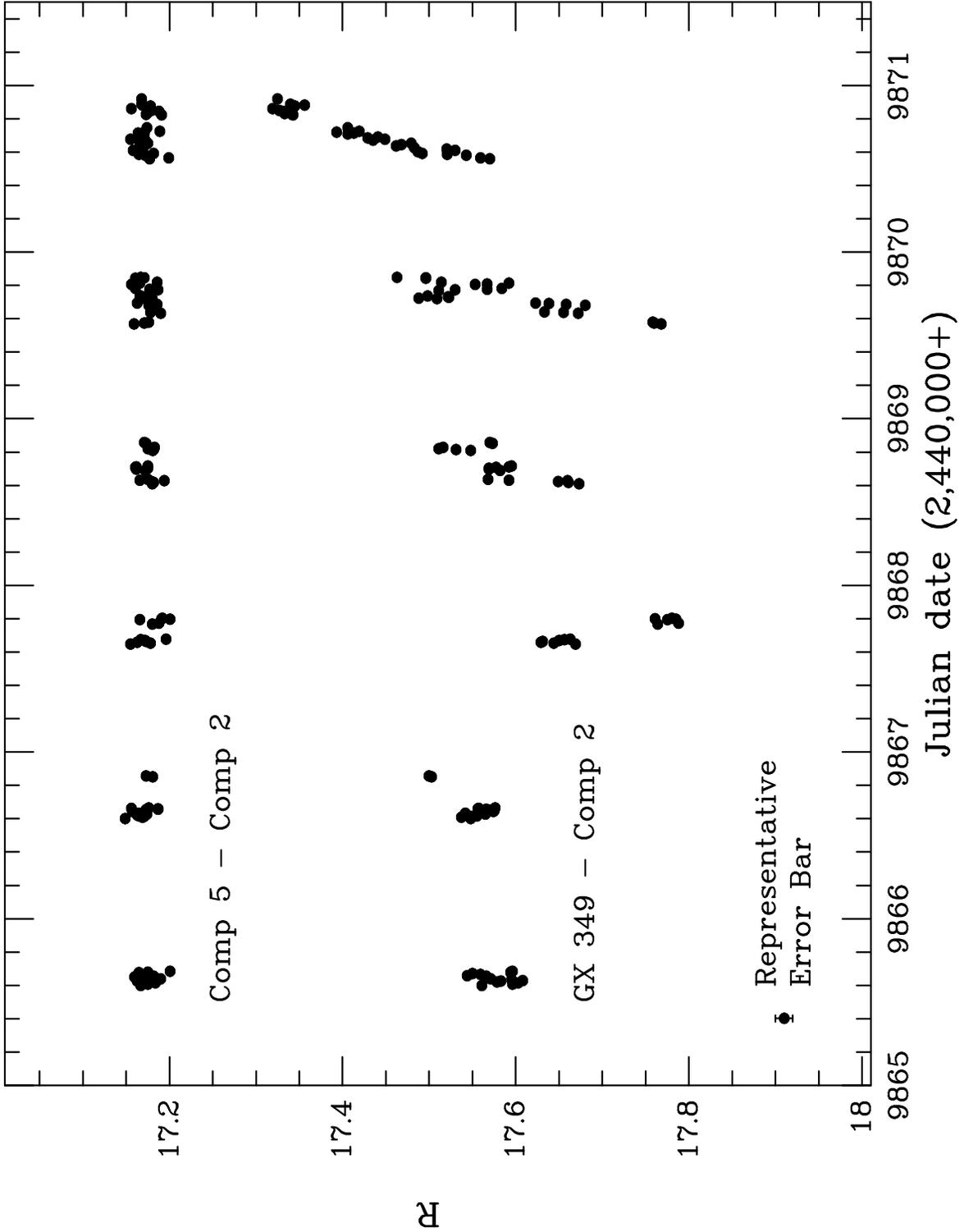}
\figcaption {Lightcurve of GX~349+2 and a comparison star of similar
brightness. The  magnitudes of comparison star 5 have been shifted to
brighter magnitude by 0.6 mag for clarity of display.  The
representative 1$\sigma$ error bar has been obtained from the rms
scatter in the differential lightcurve of the two comparison stars.
\label{f-lc}}
\end{figure}

Since previous finding charts of GX~349+2, obtained with photographic plates,
are of modest quality,   
a CCD image of the field is displayed in the top panel of
Figure~\ref{f-fc}. GX~349+2 and all comparison stars used in the analysis 
are marked.
Due to the crowded field, aperture photometry is not possible, and all
photometry was performed by point spread function fitting with DAOPHOT~II
\markcite{S}(Stetson 1993). 
The instrumental magnitudes were transformed to the standard system through
observations of several Landolt standard star fields 
(\markcite{L}Landolt 1992) obtained on two of the nights during the 
observing run. The mean
magnitudes of GX~349+2 (averaged over the whole observing run)
 and several local
comparison stars are listed in Table~\ref{t2}. The systematic error (from the 
transformation to the standard system) in these 
optical 
magnitudes is $\pm 0.04$~mag. The intrinsic 1$\sigma$ error of the relative
photometry is $\pm 0.01$~mag, derived from the rms scatter of the comparison
lightcurve (see Figure~\ref{f-lc}). 
Previous individual photometric measurements by \markcite{CP} 
Cooke \& Ponman (1991) resulted in $B=20.2$ and $V=18.7$. Their measurements 
were not taken under photometric conditions and instead stars A and B
(which had been measured by \markcite{PS}Penston et al. 1975) were used as
secondary standards.
Similar observations by \markcite{PA}Penninx \& Augusteijn (1991) 
gave $B=20.0\pm 0.1$ and
$V=18.4\pm 0.1$. Our $V$ magnitude of $18.56\pm 0.04$ agrees with 
these earlier measurements; there are no previous R band measurements with 
which to 
compare our results. 

\begin{table} 
\dummytable\label{t2} 
\end{table} 

\subsection{Near-infrared Photometry}

In addition to the optical photometry, near-infrared (IR) $JHK$ and $Ks$ 
photometry 
of GX~349+2 was performed with the Cerro
Tololo Infrared Imager (CIRIM) on the CTIO 1.5~m telescope
on 1995 June 5 and 6 UT. CIRIM uses a 
256$\times$256 HgCdTe NICMOS3 array. All observations were taken in 
the $f/13.5$ 
mode resulting in a pixel scale
of $0.65\arcsec$~pix$^{-1}$. Four 15~s exposures were coadded to obtain one  
exposure. Each observation consisted of a mosaic of nine individual frames,
with each frame shifted from the previous one by $20\arcsec$ to form a 
3$\times$3 grid
centered on the position of GX~349+2.  
Dark frames with identical integration 
times and flat field frames for each filter 
(derived from observations of an illuminated
dome spot) were obtained on each night. First, a mean dark frame was 
subtracted from all observations. Next, a sky flat was constructed
from a median of the scaled object frames and subtracted from each observation.
Finally, the data were divided by the normalized dome flat. To increase
the signal-to-noise and eliminate bad pixels,  
several of the shifted frames were aligned and combined with a bad pixel
mask. A $Ks$ frame at the same scale as the optical image is displayed
on the bottom panel of Figure~\ref{f-fc} for comparison. 
Notice the impressive increase in the brightness of star M between the $R$ and
the IR image. This star is not visible on our $V$ frame, has $R=17.2\pm0.1$,
and is saturated on our $J$, $H$ and $K$ frames. Star M was reported to
be variable and briefly considered as the optical counterpart for GX~349+2
(\markcite{Z}Zuiderwijk 1978, \markcite{G2}Glass \& Feast 1978), but 
was subsequently shown to  
probably be a Mira variable (\markcite{G1}Glass 1979).
 
Again, photometry was performed by point spread function fitting with 
DAOPHOT~II. 
The instrumental magnitudes were transformed to the standard system (CIT)
through observations of Elias faint infrared standards (\markcite{E}Elias
et al. 1982). The results for GX~349+2 and several local comparison
stars are listed in Table~\ref{t2}. The systematic errors in the IR 
photometry are 
rather large ($\pm 0.2$ mag) due to only partially photometric 
observing conditions. Furthermore, the transformation to the standard 
system was accomplished simply by a zero point offset, neglecting color 
effects.
\markcite{CP}Cooke \& Ponman (1991) obtained a low spatial resolution 
map at $H$ and
derive a magnitude of H$\approx$15. Our $H$ magnitude of $14.5 \pm 0.2$ is
probably consistent with that measurement (for which no errors are given).

\section{Results}

The $R$ band lightcurves for
GX~349+2 and a slightly fainter comparison star (star 5)
are displayed in Figure~\ref{f-lc}. The magnitude of the comparison
lightcurve has been shifted to brighter magnitudes by 0.6~mag to separate
the two curves for display purposes. 
The systematic errors from the transformation to the standard system are not
included in the error bar. 
The differential lightcurve of GX~349+2 clearly displays variability of up to 
0.3~mag during a single night. This marked intraday variability may be added
to the positional coincidence with the radio source and the observed
H$\alpha$ emission to conclude with confidence that the correct optical
counterpart to the X-ray source has been
identified. 

\begin{figure}
\plotone{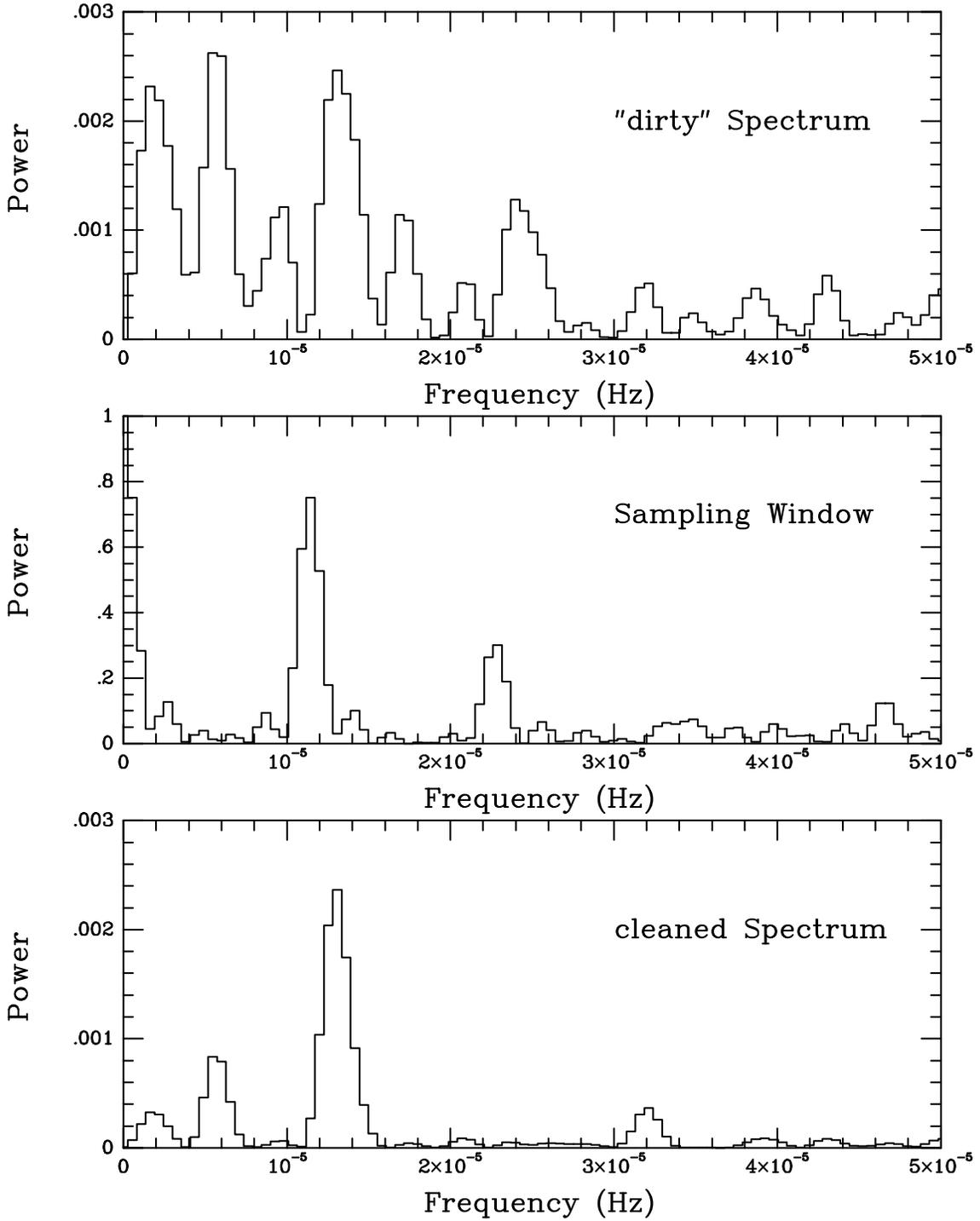}
\figcaption {Power spectrum analysis with the CLEAN algorithm.  {\it
Top panel:} the ``dirty'' spectrum, i.e., the convolution of the
periodicities in the observed data with the sampling window; {\it
middle panel:} the sampling window; {\it lower panel:} the cleaned
spectrum after 20 iterations with a gain of 0.2. The highly significant
peak at $\sim$21.4~h is clearly visible. The next strongest peak, near
50~h, is not statistically significant. \label{f-clean}}
\end{figure}

The data were searched for periodic
behavior with a power spectrum analysis using the CLEAN algorithm 
\markcite{R}(Roberts
et al. 1987). Figure~\ref{f-clean} shows the ``dirty'' spectrum, 
the power spectrum created by the sampling window, and the cleaned 
spectrum after 20 iterations of CLEAN with a gain of 0.2. The cleaned
power spectrum exhibits a strong peak around $\sim$21.4~h.
An analytic assessment of the statistical significance of the 
peak is complicated by the markedly nonuniform sampling of the data,
and the repeated application of the CLEAN algorithm. We prefer instead
a Monte Carlo simulation with boundary conditions that accurately mimic
the actual data. To this end we have scrambled the vectors of time {\it
vs.} magnitude with a random number generator, thereby creating
nonsense data with the identical sample times (and thus window
function) and standard deviation as the real data. These randomized
data were then CLEANed and transformed in the identical way as the
actual observations, and the resulting power spectrum was examined for
peaks. In 100 iterations of this scrambling process, none of the
resulting power spectra contained any peak at {\it any} frequency of
height as significant as the signal observed in the actual
data.  We conclude that the dominant peak is statistically significant
with at least 99\% confidence.
 
In order to test the reliability of the peak position, 
the dirty spectrum was cleaned with a range of gain values and
number of CLEAN iterations, and the resultant peak was
fitted with a Gaussian.
The peak center varied between 
1.29$\times10^{-5}$~s$^{-1}$ (21.53~h) and 1.31$\times10^{-5}$~s$^{-1}$
(21.20~h); the typical HWHM of the peak is 1.35~h. 
Since the power spectrum peak is so 
broad ($\pm0.9$~h lies within the width of one phase bin on either side of 
the peak center),
and it is known that CLEAN can shift the peak center frequencies in the 
presence
of noise (\markcite{Cb}Carbonell et al. 1992), we searched the period 
range 19.0~h to 23.5~h 
by folding the data on a given period in steps of 0.1~h and 
fitting a sine wave.
The period with the smallest rms after subtraction of the best
fit is 21.85~h. In order to arrive at an error estimate for this period 
we performed Monte
Carlo simulations. The best sine fit for the period of 21.85~h was 
subtracted from the data, and 
the resulting residuals scrambled and added back into the data. 
Then the process of 
folding and fitting the data was repeated. 
The resulting distribution of periods after 1000 iterations of this process
gives a 1$\sigma$ error of 0.13~h. We adopt 21.85$\pm0.4$~h for the best period 
and 3$\sigma$ error.
 
\begin{figure}
\plotone{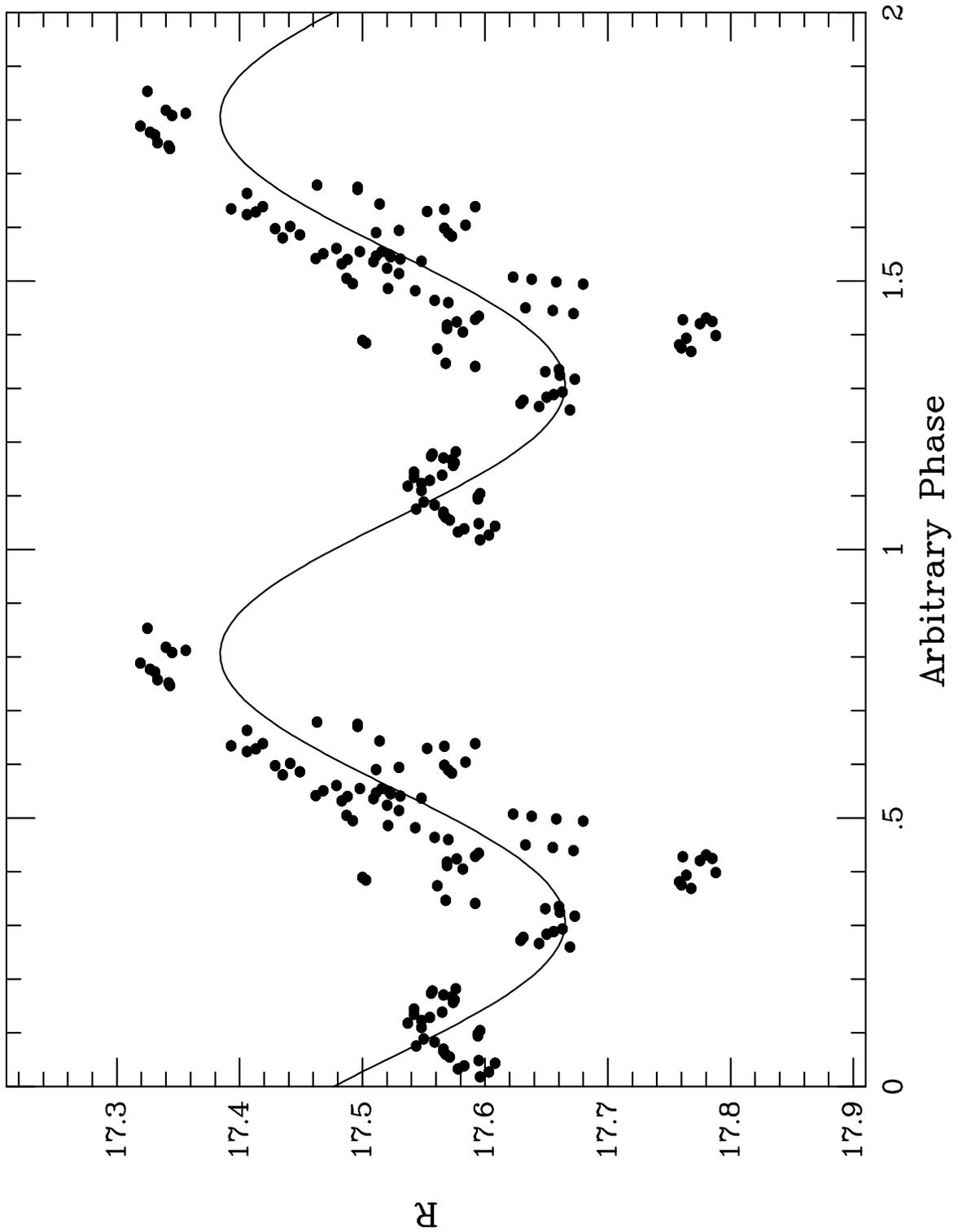}
\figcaption {The lightcurve for our photometric data of GX~349+2,
folded on the best period of 21.85~h. Also plotted is the best-fitting
sine wave.  Two cycles are shown for clarity.\label{f-fold}}
\end{figure}

Figure~\ref{f-fold} shows the photometric data folded on the best 
period of 21.85~h.
The best-fitting sine wave has been overplotted, and two cycles are shown 
for clarity. The sine fit gives a half-amplitude of 0.14~mag for the 
variations.

The folded lightcurve of GX~349+2 shows that the periodic variation is 
superposed on erratic flickering. 
This is not surprising, as both of the other 
photometrically
observed Z-sources, Sco~X-1 and Cyg~X-2, also show large scatter in their 
lightcurves \markcite{GL}(Goranskii \& Lyutyi 1988, \markcite{A}Augusteijn et 
al.\ 1992).
However, caution will be needed to insure that the variations seen in GX~349+2
are truly periodic 
and not simply due to random variability observed over a relatively 
short time span. 
Additional observations
will test the reliability of this period.    
It is interesting to note that \markcite{van}van Paradijs \& McClintock (1994) predict 
a period between 3 and 52~h for GX~349+2 from a relationship
between the absolute
visual magnitude, the X-ray luminosity, and the period of a LMXB system.

We have searched for an analog of the 21.85~h period in the X-ray emission
from GX349+2, using the preliminary public version of the All Sky
Monitor data from the Rossi X-ray Timing Explorer (quick-look results provided 
by the ASM/RXTE team, 
http://space.mit.edu/XTE/asmlc/ASM.html). 
No such periodic behavior is evident, using a variety of
analysis techniques. Although the X-ray data are well-sampled, in these
observations the source is highly and chaotically variable on a
timescale of hours, with typical amplitudes of 30\%, thereby leading to
relatively uninteresting upper limits on any periodic behavior.

Brightness variations in LMXBs can have several different
origins, e.g.,  X-ray heating of part of the surface of the secondary, 
ellipsoidal distortion of the companion  due to the gravitational influence 
of the compact object, 
a ``hot spot'' on the edge of the accretion disk, 
the accretion stream, or a combination of these effects. It is difficult to 
distinguish between these different possibilities merely on the basis of 
photometric data. For ellipsoidal variations a double 
peaked lightcurve is expected, while in the other cases a single peaked 
lightcurve is seen. As a consequence, our period could be either the orbital 
period or half the orbital period. This ambiguity can only be  
resolved 
through radial velocity measurements.

Ellipsoidal variations are usually observed in LMXB systems where the secondary
contributes significantly to the observed light, and stellar 
features of the secondary are seen in the spectrum of the system 
(e.g., Cyg~X-2, \markcite{CO}Cowley et al. 1979). No underlying stellar features
are visible in the spectrum of GX~349+2 published by \markcite{PA}Penninx and 
Augusteijn (1991).  Unfortunately, the spectrum is of rather low 
quality and those authors caution that they would have been unable to detect
stellar features of the strength seen in Cyg~X-2.
The amplitude of ellipsoidal variations is usually small, $\lesssim 0.1$~mag
full amplitude, in comparison to which the full amplitude of variations in 
GX~349+2 (0.28~mag) seems
fairly large. 
Yet, Cyg~X-2 
shows a double peaked lightcurve due to ellipsoidal variations with a full
amplitude of 0.25~mag in the $B$ band; therefore this explanation still seems
viable for GX~349+2. 

On the other hand, Sco~X-1 shows a single peaked lightcurve 
(\markcite{A}Augusteijn et al. 1992) over its 
period, and due to the similarity between Sco~X-1 and GX~349+2
in the 
X-ray regime, one might expect that the intrinsic properties of their 
optical counterparts would also resemble each other. Note that if our
period reflects the orbital period of the system, then the periods of Sco~X-1
and GX~349+2 are quite similar.   
The optical variability of Sco~X-1 is thought to
be due to X-ray heating of the disk and/or secondary. In the case of X-ray
heating of the secondary, large amplitude variations are expected over the 
orbit 
(e.g., $\sim$1.5~mag in the Her~X-1 system). 
The fact that only small amplitude variations (0.13~mag
full amplitude, \markcite{A}Augusteijn et al. 1992)
are seen in Sco~X-1 has been attributed to   
a low system inclination. If the GX~349+2 variations are also due to X-ray 
heating of the secondary,  
it might also imply a low inclination for this system. Support for 
GX~349+2 being a moderate to low inclination system comes from the fact that    
no X-ray eclipses have been reported so far.
Furthermore, \markcite{K}Kuulkers (1995)
suggested that some intrinsic differences in the X-ray properties
among the Z-sources are due to inclination effects. He groups GX~349+2
with Sco~X-1 and GX~17+2 as basically face-on systems. 
The determination of the inclination is extremely important
since this is of course one of the parameters that enters into the 
calculation of the component masses. Modeling of the photometric 
lightcurve, once
the period is confirmed and the phase coverage of the lightcurve is complete,
might be valuable in this respect.  
For the following discussion we assume that the observed variability is 
{\it not} due to ellipsoidal variations,
but that 21.85~h is the actual orbital period of the system.    

\section{Discussion}

Recently, \markcite{South}Southwell et al. (1996) suggested a period of 14~d 
for GX~349+2 from an analysis
of the H${\alpha}$ emission line velocities. However, their data are 
sampled extremely sparsely and not at the comparatively short 
timescale of our
period. There is no obvious alias relationship between the candidate 
spectroscopic and photometric periods. If the \markcite{South}Southwell et al. 
observations define the true radial velocity 
semi-amplitude of the system ($K\approx70$~km~s$^{-1}$) but our photometric 
period is the actual orbital period, then the implied mass function is 
$f(M) = 0.032$~M$_{\sun}$; 
however, both of these assumptions are currently  
uncertain. By comparison, the mass function for Sco~X-1 is
$f(M) = 0.016$~M$_{\sun}$  \markcite{CC}(Cowley \& Crampton 1975), a similarly
small value.  The period of \markcite{South}Southwell et al. (1996) would imply
$f(M)= 0.5$~M$_{\sun}$. 

\begin{figure}
\plotone{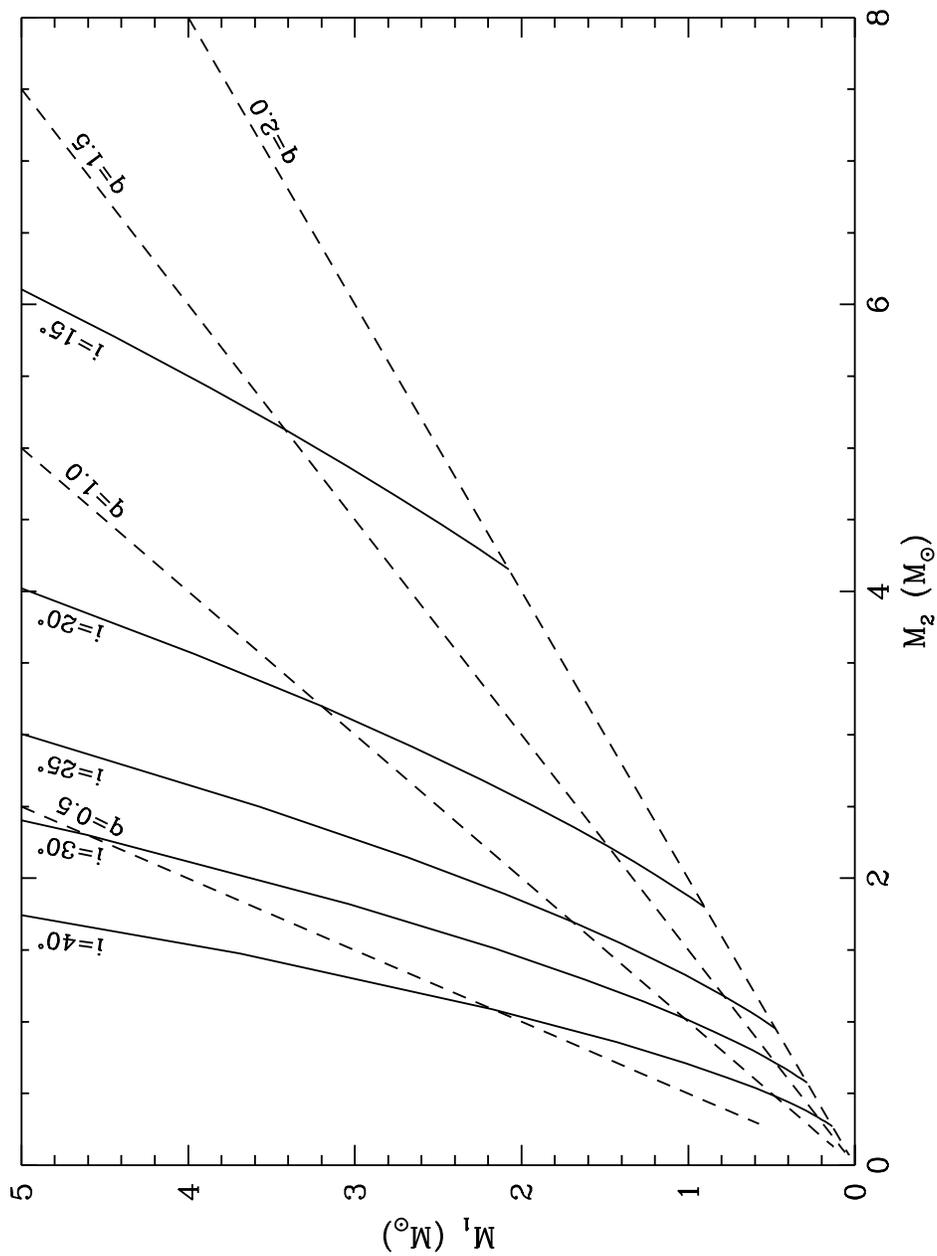}
\figcaption {Possible masses for the components in GX~349+2, adopting a
mass function $f(M) = (M_2 sin i )^3/(M_1 + M_2)^2 = 0.032$~M$_{\sun}$.
Curves of constant inclination (solid) and constant mass ratio (dashed)
are labeled.  \label{f-mf}}
\end{figure}

If the H${\alpha}$ emission arises in the
accretion disk, then the observed velocity is indicative of the neutron star; 
if instead the emission originates from the X-ray heated side of the secondary,
then 
the radial velocity traces the motion of the optical companion. This question 
can ultimately only be answered by simultaneous spectroscopy and photometry
to determine the location of the emitting region.
With our (highly uncertain) mass function of $f(M)= 0.032$~M$_{\sun}$, 
we can derive the component masses and Roche lobe 
size of the secondary for a given inclination $i$ and mass ratio $q=M_2/M_1$, 
where $M_2$ refers to the optical (mass donor) companion. 
Various curves of constant 
inclination and mass ratio are shown in  
Figure~\ref{f-mf} (calculated under the assumption that the  
H${\alpha}$ emission arises in the accretion disk). 
Assuming a mass for the neutron star of $M_1\leq 2$~M$_{\sun}$, 
we see that for $i\gtrsim 25^{\arcdeg}$ the 
mass of the companion is $M_2 < 2$~M$_{\sun}$. For $i\lesssim 15^{\arcdeg}$
the companion is more massive than the neutron star. 
We can calculate the binary separation $a$ for a given neutron star mass and
mass ratio from Kepler's law: 
\begin{equation} 
a=3.5\times10^{10}~(M_{1}/M_{\sun})^{1/3}~(1+q)^{1/3}~P_{h}^{2/3} {\rm cm}
\end{equation} 

The Roche lobe radius of the companion for a given $q$ is then obtained from the
approximate analytic formula of \markcite{Egg}Eggleton (1983)
\begin{equation}
\frac{R_L}{a} = \frac{0.49 q^{2/3}}{0.6 q^{2/3} + ln(1+q)^{1/3}}
\end{equation} 
A sample list of binary separation, Roche lobe radius and main sequence
star radius
of the corresponding $q$ is given in Table~\ref{t3}. We see that  
for up to very high mass ratios, the Roche lobe of the
secondary is too large to be filled by a main sequence star with the 
appropriate mass. Therefore, for mass transfer to occur, the companion must 
be slightly evolved. 
These conclusions are almost identical to those reached for Sco~X-1 
by \markcite{CC}Cowley \& Crampton (1975). 

\begin{table} 
\dummytable\label{t3} 
\end{table} 

We can also attempt to derive clues to the nature of the secondary from the 
photometry of GX~349+2, 
especially from the IR magnitudes. Photometry in the IR has certain 
advantages over 
optical photometry in determining the spectral type and luminosity class
for this object. First, the extinction in the IR is much smaller 
($A_K/A_V = 0.114$, \markcite{Ca}Cardelli et al. 1989), thereby reducing 
the relative size of any error
in the extinction estimate. Also, most LMXBs are 
dominated by the light from the accretion disk in the optical. 
Since the energy distribution of an accretion disk peaks in the blue,
it is expected that contamination by the disk
diminishes  towards longer wavelengths, so that the companion should be 
more easily observed. As a caveat to this statement, we note that
\markcite{CS}Cowley et al. (1991) conducted a spectroscopic
survey of LMXBs in the near-IR (8000--9000~{\AA})
and find that in most systems the accretion disk emission
still dominates the spectrum. 
Observations of cataclysmic variables showed
that the disk contribution in the IR varies from case to case, with 
several systems still indicating a strong contribution of light from the 
accretion disk (\markcite{B}Berriman et al. 1985, \markcite{F}Friend et 
al. 1988). 


The extinction for GX~349+2 has been estimated from various methods
to lie between $A_V = $4--6~mag 
(\markcite{PA}Penninx \& Augustijn 1991, \markcite{CP}Cooke \& Ponman 1991).
Adopting $A_V = 5$~mag, we arrive at dereddened magnitudes
of $J=13.6$, $H=13.55$, and $K=14.0$ for GX~349+2. Distance estimates 
for GX~349+2
range from 8.5~kpc (canonical distance to the Galactic Center, 
\markcite{CP}Cooke \& Ponman 1991)
to 10~kpc (assuming that Z-sources radiate with the Eddington 
luminosity in X-rays, 
\markcite{van}van Paradijs \& McClintock, 1994). This results in an absolute $K$ magnitude
of $M_K=-0.6$ or $M_K=-1.0$, respectively. These magnitudes 
correspond to a main sequence spectral type of roughly B5, but the 
$(J-K)_0$ color is then too blue by 0.28~mag. Although this would still be  
within our photometric errors, it seems unlikely that GX~349+2 has a 
B5 main sequence star companion. A B5 companion would account for all the 
light seen in the $B$ and $V$ band and therefore exclude the presence
of an accretion disk. This would make GX~349+2 a high mass X-ray binary 
system, which seems inconsistent with its X-ray properties. It is more
likely that the accretion disk still contributes considerable flux in the 
IR and that the IR colors are not necessarily indicative of the companion.

\section{Conclusions}

We have carried out the first photometric study of the optical counterpart of
the Z-source GX~349+2. 
We find variability of 
0.3~mag during a single night and a significant $21.85$~h period 
superposed on erratic 
flickering. Assuming this to be the orbital period of the system 
together with a 
velocity amplitude reported by \markcite{South}Southwell et al. (1996), we find a 
mass function quite 
similar to that of Sco~X-1. Infrared photometry seems to indicate that
the disk is still contributing considerable flux in the $K$ band.

\acknowledgments

We wish to thank Richard Elston for his help in reducing the 
IR data and Don Hoard
for reading a draft of this paper and 
providing helpful comments.  This research was in part supported by 
NASA grant NAG5-1630, and  
has made use of the Simbad 
database, operated at CDS, Strasbourg, France.


\clearpage

\begin{deluxetable}{ccccc}
\tablenum{1}
\tablecolumns{5}
\tablewidth{0pt}
\tablecaption{Z-sources\tablenotemark{a}}
\tablehead{
\colhead{X-ray Source} &
\colhead{Alternate Name}  &
\colhead{Optical Cpt.} &
\colhead{$V$}  &
\colhead{Period (h)}
}
\startdata
Sco X-1 & 1617$-$155 & V818 Sco & 12.2 & 18.90 \\
GX 340+0& 1642$-$455 & \nodata\tablenotemark{b}& \nodata & \nodata \\
GX 349+2& 1702$-$363 & star 6\tablenotemark{c}& 18.6 & 
          21.85\tablenotemark{d}~/334.6\tablenotemark{e}\\
GX 5-1  & 1758$-$250 & \nodata & \nodata &  \nodata \\
GX 17+2 & 1813$-$140 & NP Ser\tablenotemark{f} & 17.5 &  \nodata \\
Cyg X-2 & 2142$+$380 & V1341 Cyg & 14.7 &236.2
\enddata
\tablenotetext{a}{\markcite{v2}van Paradijs (1995)}
\tablenotetext{b}{IR counterpart suggested by \markcite{M}Miller et al. (1993)}
\tablenotetext{c}{\markcite{J}Jernigan et al. (1978)}
\tablenotetext{d}{this work}
\tablenotetext{e}{\markcite{South}Southwell et al. (1996)}
\tablenotetext{f}{identification uncertain: see \markcite{D}Deutsch et al. (1996)}
\end{deluxetable}


\begin{deluxetable}{clllll}
\tablenum{2}
\tablecolumns{6}
\tablewidth{0pt}
\tablecaption{Magnitudes of GX~349+2 and comparison stars}
\tablehead{
\colhead{Object\tablenotemark{a}} &
\colhead{$V$} &
\colhead{$R$} &
\colhead{$J$} &
\colhead{$H$} &
\colhead{$K$}
}
\startdata
GX~349+2 & 18.56 & 17.55 & 15.0 & 14.5 & 14.6 \nl
5        & 19.23 & 17.76 & \nodata\tablenotemark{b} & 13.1 & 13.1 \nl
2        & 17.99 & 16.61 & \nodata\tablenotemark{c} & \nodata\tablenotemark{c}
         & \nodata\tablenotemark{c} \nl
7        & 19.46 & 17.99 & 14.3 & 13.3 & 13.2 \nl
C1       & 19.96 & 18.34 & 14.5 & 13.4 & 13.3 \nl
\enddata
\tablenotetext{a}{nomenclature references in Fig.~1}
\tablenotetext{b}{affected by a bad pixel}
\tablenotetext{c}{too close to saturated star M}
\end{deluxetable}


\begin{deluxetable}{cccc}
\tablenum{3}
\tablecolumns{4}
\tablewidth{300.0pt}
\tablecaption{Parameters of GX~349+2}
\tablehead{
\colhead{$q$\tablenotemark{a}} &
\colhead{$a$ $(R_{\sun})$} &
\colhead{$R_L(M_2)$$(R_{\sun})$} &
\colhead{$R_{MS}(M_2)$\tablenotemark{b}  $(R_{\sun})$} 
}
\startdata
0.1  & 4.49 & 0.93 & 0.25 \nl
0.3  & 4.75 & 1.33 & 0.54 \nl
0.5  & 4.98 & 1.60 & 0.78 \nl
0.7  & 5.19 & 1.81 & 0.98 \nl
1.0  & 5.48 & 2.08 & 1.27 \nl
1.5  & 5.91 & 2.45 & 1.68 \nl
2.0  & 6.28 & 2.76 & 2.06 \nl
2.5  & 6.61 & 3.04 & 2.40 \nl
3.0  & 6.91 & 3.29 & 2.73 \nl
3.5  & 7.18 & 3.52 & 3.04 \nl
4.0  & 7.44 & 3.73 & 3.34 \nl
\enddata
\tablenotetext{a}{assuming $M_1 = 1.4$ M$_{\sun}$}
\tablenotetext{b}{assuming a $M$--$R$ relationship of $R/R_{\sun}=(M/M_{\sun})^{0.7}$}
\end{deluxetable}


\newpage

\epsscale{0.85}

\epsscale{0.80}

\epsscale{0.85}

\end{document}